\documentclass[%
twocolumn,superscriptaddress,
 amsmath,amssymb,
 aps,
prl
]{revtex4-1}

\usepackage{graphicx}
\usepackage{dcolumn}
\usepackage{bm}
\usepackage{xcolor}

\begin{document}

\preprint{APS/123-QED}

\title{Airflows inside passenger cars and implications for airborne disease transmission}

\author{Varghese Mathai}
\altaffiliation[]{Email: vmathai@umass.edu}

\affiliation{Department of Physics, University of Massachusetts, Amherst, Massachusetts 01003, USA}
\affiliation{Center for Fluid Mechanics, Brown University, Providence, RI 02912, USA}

\author{Asimanshu Das} 
\homepage{VM and AD contributed equally to this work and are joint first authors.}

\affiliation{Center for Fluid Mechanics, Brown University, Providence, RI 02912, USA}

\author{Jeffrey A. Bailey}

\affiliation{Department of Pathology and Laboratory Medicine, Warren Alpert Medical School, Brown University, Providence, RI 02912, USA}
\author{Kenneth Breuer}
\affiliation{Center for Fluid Mechanics, Brown University, Providence, RI 02912, USA}

\date{\today}

\begin{abstract}
Transmission of highly infectious respiratory diseases, including SARS-CoV-2, is often facilitated by the transport of tiny droplets and aerosols (harboring viruses, bacteria, etc.) that are breathed out by individuals and can remain suspended in air for extended periods of time, especially in confined environments. A passenger car cabin represents one such situation in which there exists an elevated risk of pathogen transmission. {\color{black} Despite its overwhelming significance in the face of a pandemic, few studies have addressed how the  in-cabin microclimate of a car can potentially spread pathogenic species between occupants}. Here we present results from numerical simulations, supported by field tests, to assess the airborne transmission risks in a passenger car, for a variety of ventilation configurations representing combinations of open and closed windows. We estimate relative concentrations and residence times of a non-interacting, passive scalar -- a proxy for infectious pathogenic particles -- that are advected and diffused by the turbulent airflows inside the cabin. An airflow pattern that travels across the cabin, entering and existing farthest from the occupants can potentially reduce the transmission. {\color{black} Our findings reveal the complex fluid dynamics at play during everyday commutes, and non-intuitive ways in which open windows enhance or suppress the airborne transmission pathways.}
\end{abstract}

\maketitle


\noindent {\color{black}Outbreaks of respiratory diseases, such as influenza, SARS, MERS and now the novel coronavirus~(SARS‑CoV‑2), have taken a significant toll on human populations worldwide}. They are redefining a myriad of social and physical interactions as we seek to control the predominantly airborne transmission of the causative, severe acute respiratory syndrome coronavirus disease-2 \cite{Morawska2020,zhang2020identifying,yu2004evidence}. One common and critical social interaction that must be reconsidered is how people travel in passenger automobiles, as driving in an enclosed car cabin with a co-passenger can present a significant risk of airborne disease transmission. {Most megacities (e.g. New York City) support over a million of such rides every day with median figures of 10 daily interactions per rider \cite{smith2020superspreaders}.} For maximum social isolation, driving alone is clearly ideal but this is not widely practical or environmentally sustainable, and there are many situations in which two or more people need to drive together.  Wearing face masks and using  of barrier shields to separate occupants do offer an effective first step towards reducing infection rates \cite{Tang2011,Lai2012,Greenhalgh2020,Leung2020,lee2008respiratory,plasticpanel2020roshun}. However, aerosols can pass through all but the most high-performance filters \cite{Leung2020,mittal2020flow} and virus emissions via micron-sized aerosols associated with breathing and talking, let alone coughing and sneezing, are practically unavoidable \cite{Gupta2010,bourouiba2020turbulent,Meselson2020,yan2018infectious,wolfel2020virological,yang2011concentrations,scharfman2016visualization,bahl2020airborne,bourouiba2014violent}. {\color{black} Even with basic protective measures such as mask-wearing, the in-cabin micro-climate during these rides falls short on a variety of epidemiological guidelines \cite{liu2020aerodynamic} with regard to occupant-occupant separation and interaction duration for a confined space.}  Preliminary models indicate a build-up of the viral load inside a car cabin for drives as short as 15 minutes \cite{somsen2020small,allen2020coronacar}, with evidence of virus viability within aerosols of up to 3 hours \cite{van2020aerosol,stadnytskyi2020airborne}.



To assess these risks, it is critical to understand the complex airflow patterns that exist inside the passenger cabin of an automobile, and furthermore, to quantify the air that might be exchanged between a {\it driver} and a {\it passenger}. Although the danger of transmission while traveling in a car has been recognized \cite{Knibbs2012}, published investigations of the detailed air flow inside the passenger cabin of an automobile are surprisingly sparse.
Several works have addressed the flow patterns inside automobile cabins, but only in the all-windows-closed configuration \cite{Alexandrov2001,Lee2011,Ullrich2018} -- most commonly employed  so as to reduce noise in the cabin.  At the same time, an intuitive means to minimize infection pathways is to drive with some or all of the windows open, presumably enhancing the fresh air circulating through the cabin. 

Motivated by the influence of pollutants on the passengers, a few studies have evaluated the concentration of contaminants entering from outside the cabin \cite{Muller2011} and the persistence of cigarette smoke inside the cabin subject to different ventilation scenarios \cite{Ott2008,Saber2011}.  However, none of these studies have addressed the \emph{micro-climate} of the passenger enclosure, and the transport of a contaminant from a specific person in the cabin (e.g. the  driver) to another specific person (e.g. a passenger). In addition to this being an important problem applicable to airborne pathogens in general, given that the COVID-19 pandemic is likely to present a public health risk for several months or years to come, the need for a quantitative assessment of such airflow patterns inside the passenger cabin of an automobile seems urgent. 

The current work presents a quantitative approach to this problem. While the range of car geometries and driving conditions is vast, we restrict our attention to that of two people driving in a car, which is close to the average occupancy in passenger cars in the United States \cite{Averageoccupancy2019}. 
We ask the question: what is the transport of air and potentially infectious aerosol droplets between the {\it driver} and the {\it passenger}, and how does that air exchange change for various combinations of open- and closed windows. 

\begin{figure}
\centering
\includegraphics[width=0.5\textwidth]{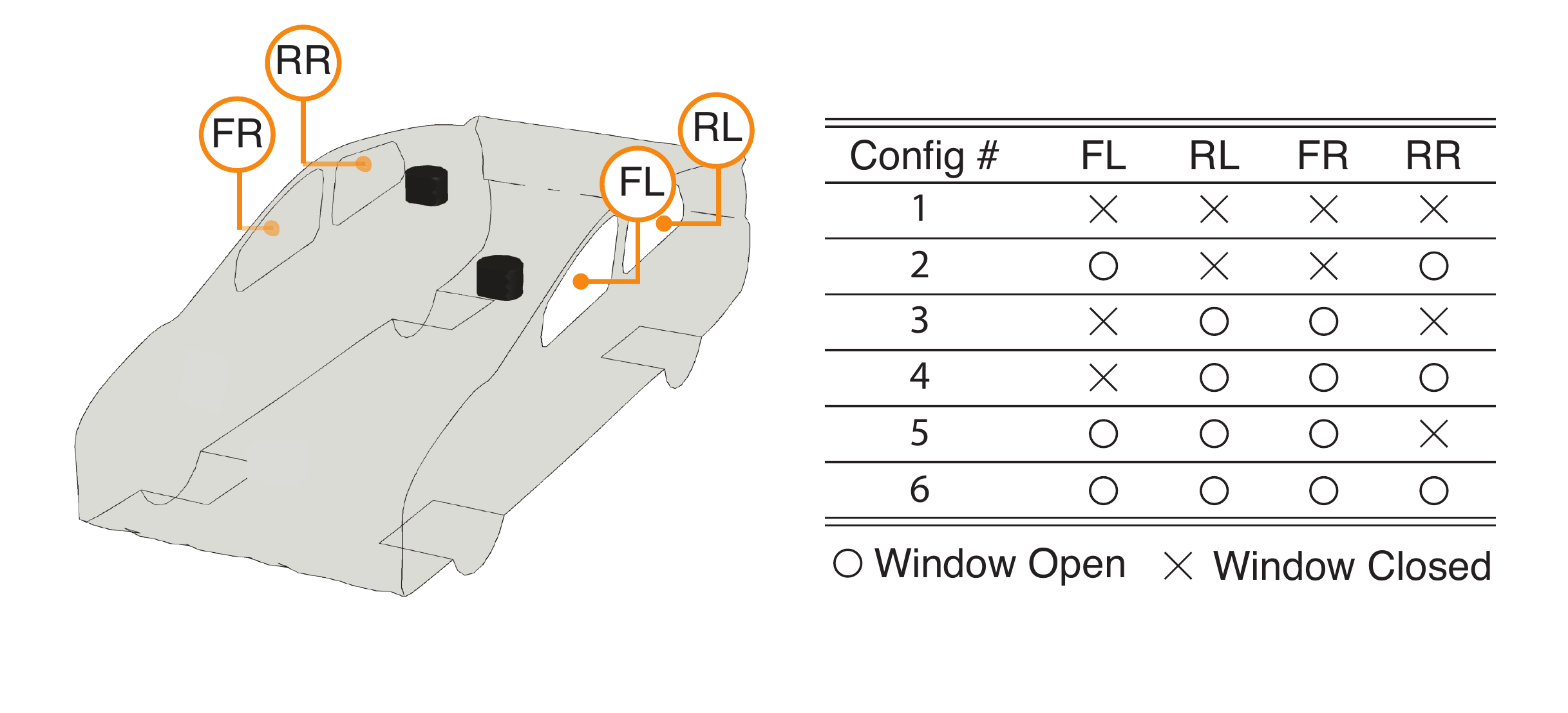}
\caption{Schematic of the model car geometry, with identifiers the front-left~(FL), rear-left~(FL), front-right~(FL), and rear-right~(FL) windows. The two regions colored in black represent the faces of the {\it driver} and the {\it passenger}. Table on the right summarizes the six configurations simulated, with various combinations of open- and closed windows.}
\label{fig:Car_configuration}
\end{figure}

To address this question, we conducted a series of representative Computational Fluid Dynamics (CFD) simulations for a range of ventilation options in a model four-door passenger car.  The exterior geometry was based on a Toyota Prius, and we simulated the flow patterns associated with the moving car while having a hollow passenger cabin and six combinations of open and closed windows, named as front-left~(FL), rear-left (RL), front-right (FR) and rear-right (RR) (Fig.~\ref{fig:Car_configuration}). We consider the case of two persons traveling in the car -- the {\it driver} in the front left-hand  seat (assuming a left-hand-drive vehicle) and the {\it passenger} sitting in the rear right-hand seat, thereby maximizing the physical distance~($\approx 1.5$ m) between the occupants. For the purposes of simulation, the occupants were modeled simply as cylinders positioned in the car interior.  

As a reference configuration (Fig. 1, Config. 1), we consider driving with all four windows closed and a typical air-conditioning flow -- with air intake at the dashboard and outlets located at the rear of the car -- that is common to many modern automobiles \cite{khatoon2020thermal}. 
The intake air was modeled to be fresh (i.e. no re-circulation) with a relatively high inflow rate of 0.1 kg/s \cite{fojtlin2016airflow}. 

The numerical simulations were performed using ANSYS-Fluent package, solving the three-dimensional, steady, Reynolds-averaged Navier-Stokes (RANS) equations using a standard $k-\epsilon$ turbulence model (for details see Methods section). We simulated a single driving speed of $v =$ 22 m/s (50 mph) and an air  density, $\rho_a$ = 1.2 kg/m$^3$. This translates to a Reynolds number of 2~million~(based on the car height), which is high enough that the results presented here should be insensitive to the vehicle speed. The flow patterns calculated for each configuration were used to estimate the air (and potential pathogen) transmission from the {\it driver} to the {\it passenger}, and conversely from the {\it passenger} to the {\it driver}. These estimates were achieved by computing the concentration field of a passive tracer ``released'' from each of the occupants, and evaluating the amount of that tracer reaching the other occupant (see Methods section).

In this paper, we first describe the pressure distributions established by the car motion and the flow induced inside the passenger compartment. Following that we describe the passenger-to-driver and driver-to-passenger transmission results for each of the ventilation options, and finally conclude with insights based on the observed concentration fields, and general conclusions and implications of the results.

\begin{figure} [!b]
\centering
\includegraphics[width=0.5\textwidth]{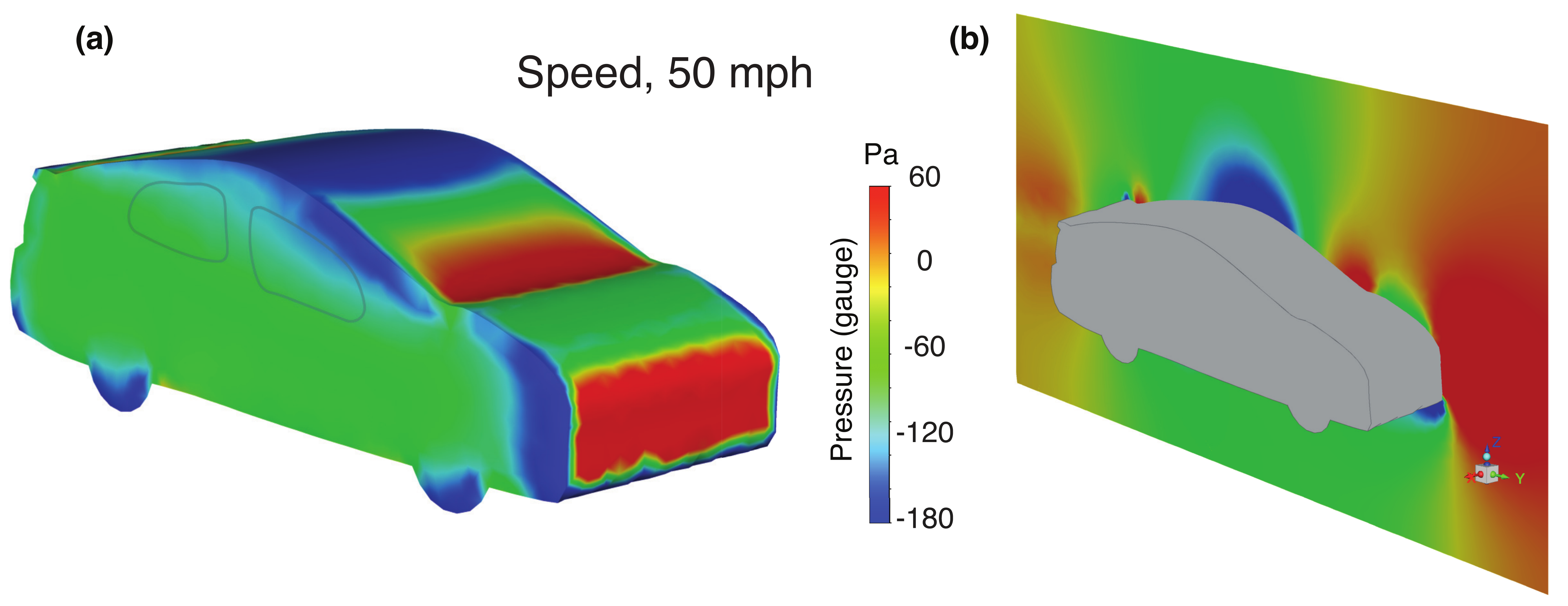}
\caption{Pressure distributions around the exterior of the car, associated with a vehicle speed of 22 m/s (50 mph). (a) Surface pressure distribution.  (b) Pressure distribution in the air at the mid-plane. The color bar shows the gauge pressure in Pascal, and emphasizes the mid-range of pressures: $[-180, 60]$ Pa; at this speed the full range of gauge pressure on the surface is $[-361, 301]$ Pa.}
\label{fig:pressure}
\end{figure}

\section*{Results and Discussion}

\subsection*{Overall air flow patterns}
The external airflow generates a pressure distribution over the car  (Fig.~\ref{fig:pressure}), forming a high-pressure stagnation region over the radiator grille and on the front of the windshield. The peak pressure here (301 Pa) is of the order of the dynamic pressure ($0.5 \rho_a v^2  = 290$ Pa at 22 m/s). Conversely, as the airflow  wraps over the top of the the car and around the sides, the high air speed is associated with a low pressure zone, with the local pressure well below atmospheric (zero gauge pressure in Fig~\ref{fig:pressure}). This overall pressure map is consistent with other computations of flows over automobile bodies \cite{parab2014aerodynamic} and gives a physical preview to a key feature -- that the areas near the front windows and roof of the car are associated with lower-than-atmospheric pressures, while the areas towards the rear of the passenger cabin are associated with neutral or higher-than-atmospheric pressures.

\begin{figure}
\centering
\includegraphics[width=0.4\textwidth]{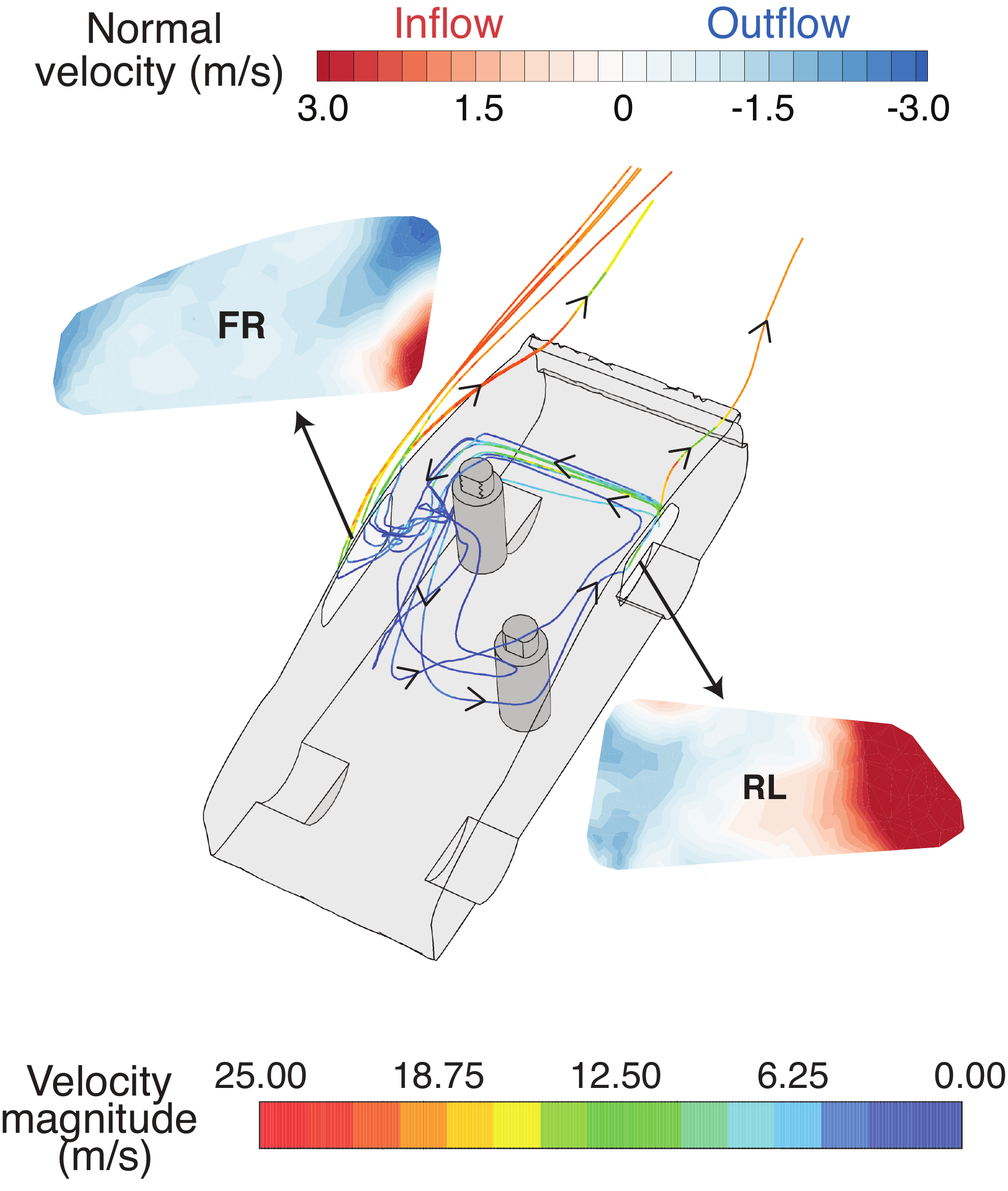}
\caption{Streamlines computed for the case in which the rear-left and front-right windows are open.  The streamlines were initiated at the rear-left (RL) window opening. The streamline color indicates the flow velocity. Insets show the front-right (FR) and RL windows colored by the normal velocity. The RL window has a strong inflow (positive) of ambient air, concentrated at its rear, whereas the front right window predominantly shows an outward flow (negative) to the ambient.}
\label{fig:RL-FR-streamline}
\end{figure}

A typical streamline (or pathline) pattern in the car interior is shown in Fig.~\ref{fig:RL-FR-streamline}, where the rear-left and front-right windows are opened~(Config. 3 in Fig.~\ref{fig:Car_configuration}).  The streamlines were initiated at the rear-left (RL) window which is the location of a strong inflow (Fig.~\ref{fig:RL-FR-streamline}-lower right),due to the high pressure zone established by the car's motion (Fig~\ref{fig:pressure}).  A strong air current ($\sim 10$ m/s) enters the cabin from this region and travels along the back seat of the car, before flowing past the {\it passenger} sitting on the rear-right side of the cabin.  The air current turns at the closed rear-right window, moves forward and the majority of the air exits the cabin at the open window on the front-right (FR) side of the vehicle, where the exterior pressure is lower than atmospheric (Fig~\ref{fig:pressure}). There is a much weaker air current ($\sim 2$ m/s) that, after turning around the {\it passenger}, continues to circulate within the cabin. A small fraction of this flow is seen to exit through the RL window.

The streamline arrows indicate that the predominant direction of the recirculation zone inside the cabin is counter-clockwise (viewed from above). These streamlines, of course, represent possible paths of transmission, potentially transporting virus-laden droplets or aerosols throughout the cabin and, in particular, from the {\it passenger} to the {\it driver}. 

\begin{figure}[!b]
\centering
\includegraphics[width=1.8in]{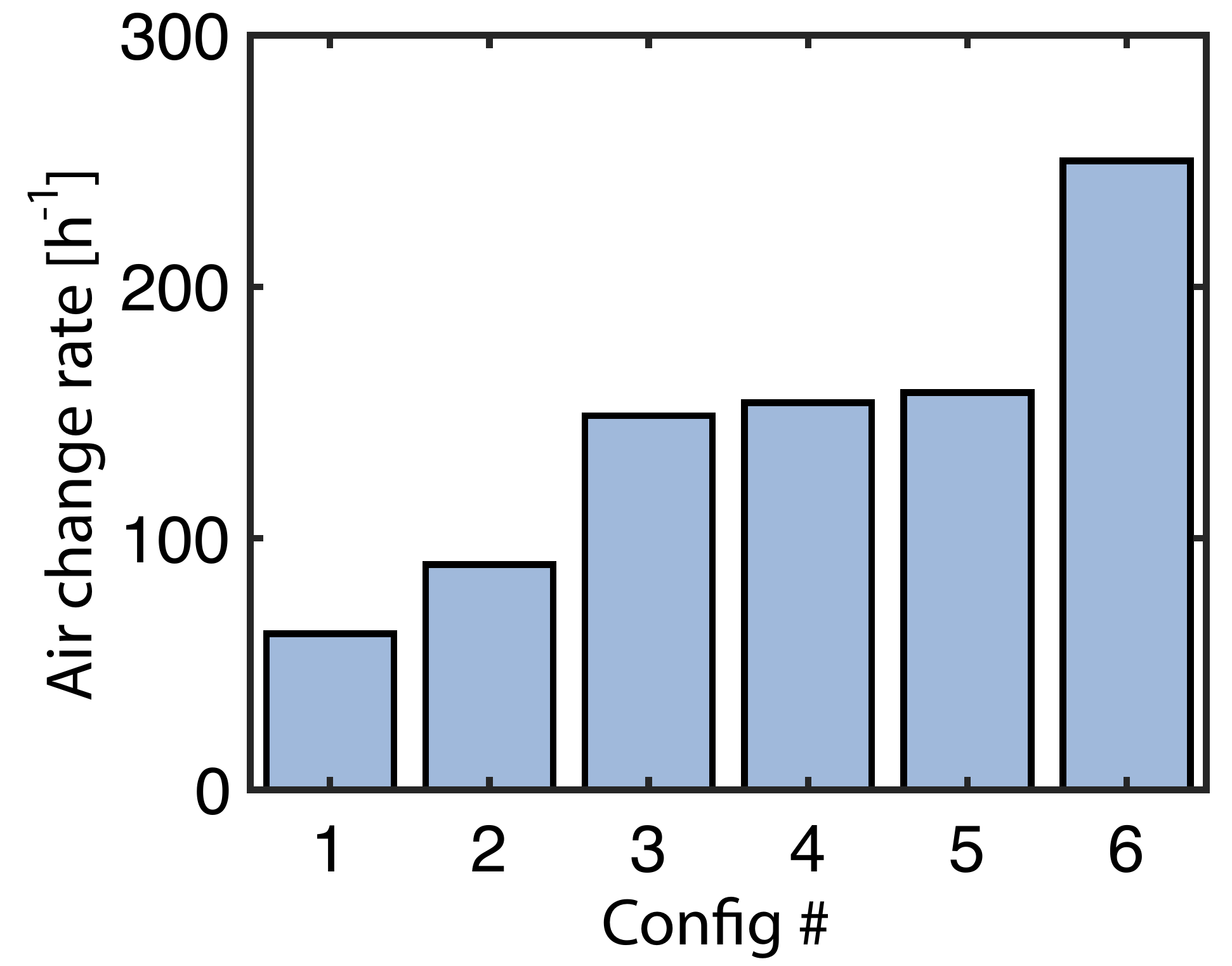}
\caption{Air change rate (or ACH) calculated based on a residence time analysis for different configurations. Here, the air change rate per hour is given by $1/\tau_r$, where $\tau_r$ is the residence time in hours.}
\label{fig:ACH}
\end{figure}

As already indicated, for the particular ventilation option shown here, the overall air pattern -- entering on the rear-left and leaving on the front-right -- is consistent with the external pressure distributions (Fig~\ref{fig:pressure}). The elevated pressure towards the rear of the cabin and the suction pressure near the front of the cabin drive the cabin flow. 
This particular airflow pattern was confirmed in a ``field test'' in which the windows of a test vehicle (2011 Kia Forte hatchback) were arranged as in Config. 3, i.e.  with the RL and FR windows open. The car was driven at 30 mph on a length of straight road, {\color{black} a flow wand (a short stick with a cotton thread attached to the tip) and a smoke generator were used to visualize the direction and approximate strength of the air flow throughout the cabin.} By moving the wand and the smoke generator to different locations within the cabin, the overall flow patterns obtained from the CFD simulations -- a strong air stream along the back of the cabin that exits the front-right window, and a very weak flow near the {\it driver} -- were qualitatively confirmed (see supplemental materials Figs.~S-1, S-2 and S-3). Different ventilation configurations generate different streamline patterns (e.g. Supplementary materials Figs.~S-4 and S-5) but can all be linked to the pressure distributions established over the car body (Fig~\ref{fig:pressure}).

An important consideration when evaluating different ventilation options in the confined cabin of a car is the rate at which the cabin air gets replenished with fresh air from the outside. This was measured by Ott et al.~\cite{Ott2008} for a variety of cars, traveling at a range of speeds, and for a limited set of ventilation options. In these measurements, a passive tracer~(representing cigarette smoke) was released inside the cabin and the exponential decay of the tracer concentration measured. Assuming the cabin air to be well-mixed \cite{Ott2008}, they estimated the air-changes-per-hour (ACH) -- a widely used metric in indoor ventilation designs.

\begin{figure*}[!t]
\centering
\includegraphics[width=.85\textwidth]{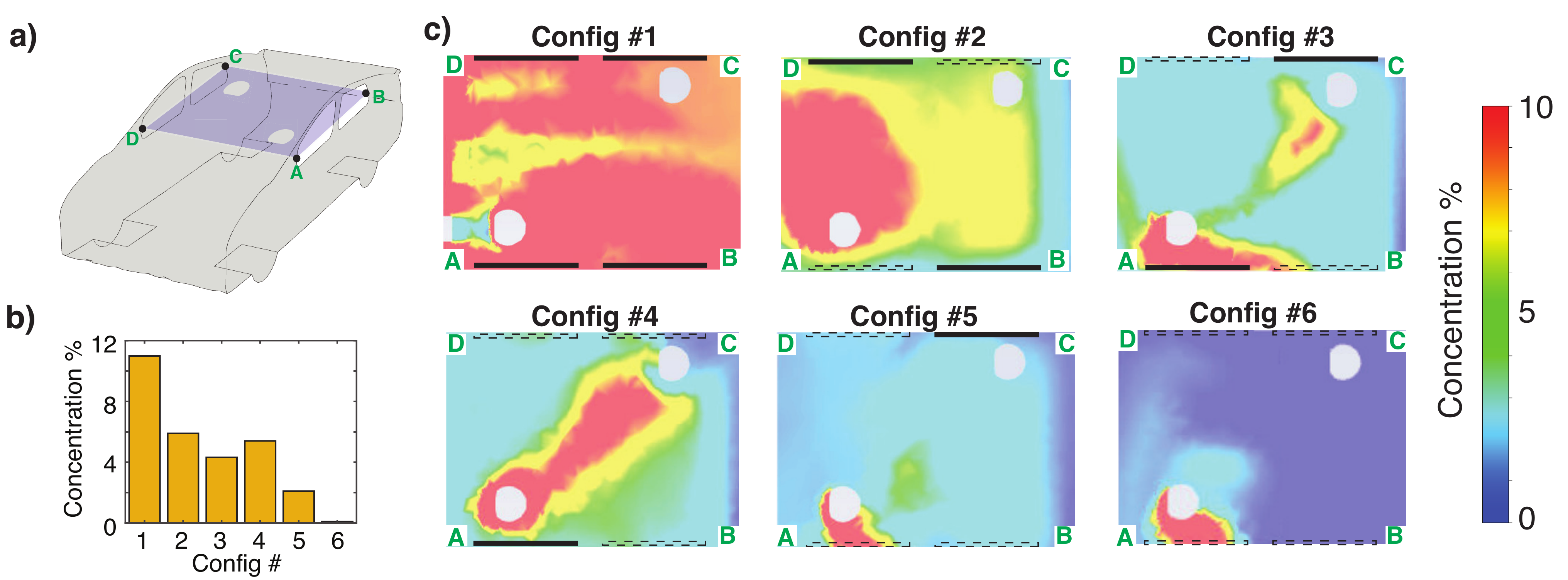}
\caption{Driver-to-passenger transmission. (a) Schematic of the vehicle with a cut plane passing through the center of the inner compartment on which the subsequent concentration fields are shown. (b) The bar-graph shows the mass fraction of air reaching the {\it passenger} that originates from the {\it driver}. (c) shows the concentration field of the species originating from the {\it driver} for different window cases. The dotted and the solid lines denote the open and closed windows, respectively. Note that the line segment A-D is at the front of the car cabin, and the flow direction in (c) is from left to right. Here the dotted line represent open window and solid line indicate closed window.}
\label{fig:drivertopassenger}
\end{figure*}


From the simulations, we can precisely compute the total flow of air entering (and leaving) the cabin and, knowing the cabin volume, we can compute the air-changes-per-hour directly. Such a calculation yields a very high estimate of ACH (of the order of 1000, see Supplemental Fig.~S-6), but this is misleading, since the assumption of well-mixed cabin air is an over-simplification. Instead, a more relevant quantification of the ACH was obtained using a residence time analysis (RTA) for a passive scalar released at multiple locations within the passenger cabin. The time taken for the concentration at the outlets to decay below a threshold (1\% of the initial value) was computed, and the inverse of this time yields effective values for ACH (Fig.~\ref{fig:ACH}) which compare favorably with those reported by Ott et al. \cite{Ott2008}, after correcting for the vehicle speed  \cite{fletcher1994air}.

As one might expect, configuration 6 -- all windows open -- has the highest ACH  - approximately 250, while among the remaining configurations, Configuration 1  -- all windows closed -- has the lowest ACH of 62. However, what is somewhat surprising is that the ACH for Config. 2, in which the the windows adjacent to the {\it driver} and the {\it passenger} (FL and RR, respectively) are opened is only 89 - barely higher than the all-windows-closed configuration; the remaining three configurations (Configs. 3, 4 \& 5) with two or three open windows all show relatively high efficacy of about 150 ACH. The reason for these differences can be traced back to the overall streamline patterns and the pressure distributions that drive the cabin flow (Fig.\ref{fig:pressure}). A well-ventilated space requires the availability of an entrance and an exit, and a favorable pressure gradient between the two. Once such a cross-ventilation path is established (as in Config. 3 or Fig.~\ref{fig:RL-FR-streamline}), opening a third window has little effect on the ACH. However, we will later show that the ACH gives only a partial picture and the spreading of a passive scalar can show marked variations between the configurations 3--5, despite their nearly constant ACH.



\subsection*{Driver-to-Passenger transmission}\label{DtoP}
The flows established through the cabin provide a path for air transmission between the two occupants, and hence a possible infection route. Our focus here is on transmission via aerosols, which are small enough (and non-inertial) that they can be regarded as faithful tracers of the fluid flow \cite{mathai2016microbubbles,warhaft2000passive}.

We begin by addressing the problem from the viewpoint of an infected {\it driver} releasing pathogen-laden aerosols and potentially infecting the {\it passenger}. Fig.~\ref{fig:drivertopassenger} shows a comparison of the spreading  patterns of a passive scalar released near the {\it driver} and reaching the {\it passenger} (for details, see Methods section). To obtain a volumetric quantification, the average scalar concentration in a 0.1 m diameter spherical domain surrounding the {\it passenger's} face is also computed, as shown in Fig.~\ref{fig:drivertopassenger}(b).

\begin{figure*}[!t]
\centering
\includegraphics[width=0.85\textwidth]{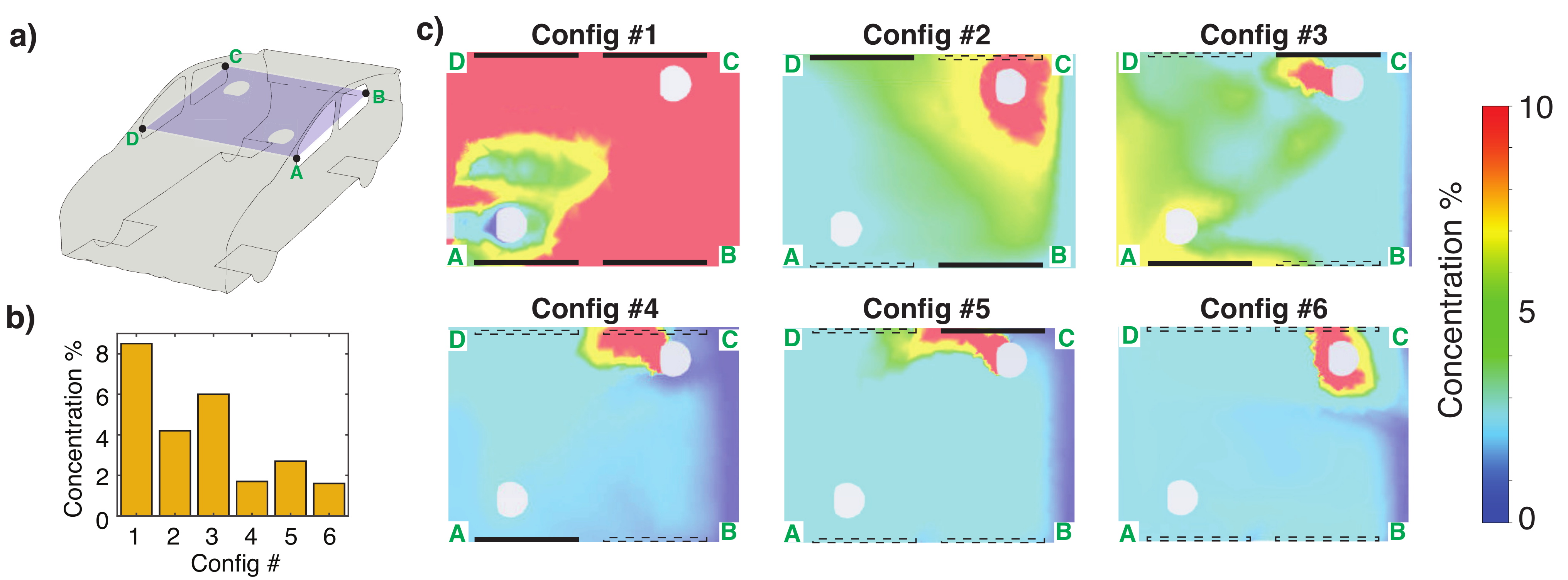}
\caption{Passenger-to-driver transmission. (a) Schematic of the vehicle with a cut plane passing through the center of the inner compartment on which the subsequent concentration fields are shown. (b) The bar-graph shows the mass fraction of air reaching the {\it driver} that originates from the {\it passenger}. (c) shows the concentration field of the species originating from the  {\it passenger} for different window configuration. Here the dotted line represent open window and solid line indicate closed window.}
\label{fig:passengerto driver}
\end{figure*}

The all-windows-closed configuration fares the worst and results in over 10\% of the scalar that leaves the {\it driver}  reaching the {\it passenger}. In contrast, the all-windows-open setting (Config. 6) appears to be the best case, with almost no injected scalar reaching the {\it passenger}. An overall trend of decreasing transmission is observed when the number of open windows are increased. However, there is some variability between the different configurations, the reasons for which may not be clear until one looks at the overall flow patterns (e.g. Fig.~\ref{fig:RL-FR-streamline}).

Figure~\ref{fig:drivertopassenger}(c) shows the concentration field of the scalar in a horizontal plane A-B-C-D within the car cabin roughly at head height of the occupants (Fig.~\ref{fig:drivertopassenger}(a)). The scalar field concentration is the highest for Config. 1, where all four windows are closed. We note that this driving configuration might also represent the most widely preferred one in the United States (with some seasonal variations). Config. 2 represents a two-windows open situation, wherein the {\it driver} and the {\it passenger} open their respective windows. One might assume that this is a logical thing to do for avoiding infection from the other occupant. Although this configuration does improve over the all-windows closed situation, shown in  Fig.~\ref{fig:drivertopassenger}(b), one can see from the concentration field that Config. 2 does not effectively dilute the tracer particles, and the {\it passenger} receives a fairly large contaminant load from the {\it driver}. To explain this result, we looked more closely at the air flow patterns.  In analogy with the streamlines associated with Config 3 (Fig.~\ref{fig:RL-FR-streamline}), Config 2 establishes a strong air current from the open rear window (RR) towards the open front window (FL), and a clockwise recirculating flow within the cabin (viewed from above). Although this flow pattern is weak,  it increases the transport of  tracer  from the {\it driver} to the {\it passenger}. Moreover, the incoming air stream in Config. 2 enters behind the {\it passenger} and is ineffective in flushing out  potential contaminants emanating from the {\it driver}.

An improvement to this configuration can be achieved if two modifications are possible: i) a change in the direction of the internal circulation, and ii) a modified incoming air flow that impinges the {\it passenger} before leaving through the open window on the front. This has been realized in the two-windows-open Config. 3 (Fig.~\ref{fig:drivertopassenger}(c)), wherein the RL and FR windows are open (same as the configuration shown in Fig.~\ref{fig:RL-FR-streamline}). Now, the incoming clean air stream from the RL window partially impinges on the {\it passenger} (seated in the RR seat) as it turns around the corner. This stream of air might also act as a ``air curtain'' \cite{foster2007three}, and hence the concentration of potentially contaminated air reaching the {\it passenger} is reduced.

The remaining configurations (Configs. 4--6) will be treated as modifications made to Config. 3, by opening more windows. Config. 4 has three windows open (Fig.~\ref{fig:drivertopassenger}(c)). Since this represents opening an additional window (RR), it may be surprising to find a detrimental effect on the concentration field and the ACH (comparing Configs. 3 \& 4 in Fig.~\ref{fig:drivertopassenger}(b),(c)). The increase in the concentration can be linked to the modified air flow patterns that result from opening the third (RR) window. Firstly, opening the RR window leads to a reduction in the flow turning at the rear-right end of cabin, since a fraction of the incoming air gets bled out of this window (Fig.~S-4). Due to this diversion of the air flow, the region surrounding {\it passenger}  is less effective as a barrier to the scalar released by the {\it driver}. Secondly, the modified flow also creates an entrainment current from the {\it driver} to the {\it passenger}, which further elevates the scalar transport. 

Config. 5 presents the scenario where the third open window is the FL. This modification leads to an improvement, nearly halving the average concentration when compared to that in Config. 3. The reason for this is apparent from the concentration field (Fig.~\ref{fig:drivertopassenger}(c)). Now that the FL window near the {\it driver} is open, the relatively low pressure near the front of the car creates an outward flow that flushes out much of released species. With the substantially reduced initial concentration field near the {\it driver}, the fraction reaching the {\it passenger} is proportionately reduced. Thus, among the configurations with three windows open, Config. 5 might provide the best benefit from the viewpoint of driver-to-passenger transmission. 

Lastly, when all four windows are opened (Config. 6),  we can again use the exterior pressure distribution to predict the flow directions. The streamlines enter through the rear windows (RL and RR) and leave via the front windows (FL and FR). However, unlike the configuration with only two windows open (Fig.~\ref{fig:RL-FR-streamline}), the overall flow pattern is substantially modified (Fig.~S-5) and the streamlines obey left-right symmetry and, for the most part, do not cross  the vertical mid-plane of the car.
In this configuration, the flow is largely partitioned into two zones 
creating two cross-ventilation paths in which the total air flow rate is nearly doubled when compared to the two and three window open configurations (Fig.~S-6).

\subsection*{Passenger-to-Driver transmission} \label{PtoD}

In this section, we look into the particle (and potential pathogen) transmission from the  {\it passenger} to the {\it driver}. Fig.~\ref{fig:passengerto driver} shows a comparison of the spreading  patterns of a passive scalar within the car cabin. 
The general trend suggests a decreasing level of transmission as the number of open windows is increased, similar to the results found for the driver-to-passenger transmission. Config. 1 (all windows closed) shows the highest concentration level at the {\it driver} ($\sim 8$\%). This value, however, is lower than the 11\% reported for the inverse transport, i.e. from the {\it driver} to the  {\it passenger} (Fig.~\ref{fig:drivertopassenger}(b)), a difference that can be attributed to the fact that the air-conditioning creates a front-to-back mean flow. 

As before, the lowest level of scalar transport corresponds to Config. 6 with all windows open, 
although we note that the concentration load here (about 2\%), is noticeably higher than that for the driver-to-passenger transmission (about 0.2\%). The streamline patterns for this configuration (Supplemental Fig.~S-5) show that the air enters through the rear windows,(RL and RR) and exits through the respective front windows (FL and FR). There is, therefore, an average rear-to-front flow in both the left and right halves of the cabin which enhances transmission from the  {\it passenger} to the {\it driver}.

Among the remaining configurations (Configs. 2--5), Config. 3 shows a slightly elevated level of average concentration. The counter-clockwise interior circulation pattern is at the heart of this transmission pattern. A substantial reduction in the average concentration can be achieved by additionally opening the rear window adjacent to the  {\it passenger} (Config. 4). This allows for much of the scalar released by the  {\it passenger} to be immediately flushed out through the rear window, analogous to the way in which opening the driver-adjacent (FL) window  helps to flush out the high concentration contaminants from the {\it driver} before they can circulate to the  {\it passenger} (Fig.~\ref{fig:drivertopassenger}(c) -- Config.~5).


\section*{Concluding remarks}\label{concl} 

In summary, the flow patterns and the scalar concentration fields obtained from the CFD simulations demonstrate that establishing a dominant cross-ventilation flow within the car cabin is crucial to the minimization of particle transport between car occupants and to lessening the exchange of potentially infectious micro-particles. With this flow pattern established, the relative positions of the {\it driver} and  {\it passenger} determine the quantity of of air transmitted between the occupants.

It is, perhaps, not surprising that the most effective way to minimize cross-contamination between the occupants is to have all of the windows open (Config. 6). This establishes two distinct air flow paths within the car cabin which help to isolate the left and right sides, 
and maximizes the ACH in the passenger cabin. Nevertheless, driving with all windows open might not always be a viable or desirable option, and in these situations there are some non-intuitive results that are revealed by the calculations.

The all-windows-closed scenario with air-conditioning on (Config. 1) appears to be the least effective option. Perhaps most surprising is that an intuitive option -- of opening the windows adjacent to each occupant (Config. 2) is effective, but not always the best amongst the partial ventilation options. Config. 3, in which the two windows farthest from the occupants (FR and RL, respectively) are open, appears to give better protection to the {\it passenger}. The particular airflow patterns that the pressure distributions establish -- channeling fresh air across the rear seat, and out the front-right window --  help to minimize the interaction with the {\it driver} in the front left position. 

The role of car speed cannot be ignored when addressing the transport between the vehicle's occupants. Since the Reynolds number of the flow is high, the air flow \emph{patterns} will be largely insensitive to how fast the car is driven. However, the air-changes-per-hour (ACH) is expected to depend linearly on the car speed \cite{fletcher1994air} and consequently, the slower the car speed, the lower the ACH,  the longer the residence time in the cabin, and hence the higher the opportunity for pathogenic infection. 

The findings reported here can be translated to right-hand-drive vehicles, of relevance to countries like the UK and India.  In those situations similar, but mirrored flow patterns can be expected.  Furthermore, although the computations were performed for a particular vehicle design (loosely modeled on a Toyota Prius), we expect the overall conclusions to be valid for most four-windowed passenger vehicles. However, trucks, minivans and cars with an open moon-roof could exhibit different airflow patterns and hence different scalar transport trends.

There are, to be sure, uncertainties and limitations in our analyses approach. The simulations solve for a steady turbulent flow, while the transmission of scalar particles that might represent pathogenic aerosols will be affected by large scale, unsteady, turbulent fluctuations, which are fully captured in the present work. These effects could change the amount of tracer emitted by one occupant and reaching the other \cite{dimotakis2005turbulent}. Furthermore, although RANS simulations represent a widely-used model for scientific, industrial and automotive applications \cite{bardina1997turbulence}, there are known limitations to its predictive capabilities, and a more accurate assessment of the flow patterns and the turbulent dispersion requires a much more computationally intensive set of simulations, such as Large Eddy simulation (LES) or Direct Numerical Simulations (DNS)  -- an undertaking far beyond the aims and scope of this work. Nevertheless, despite these caveats, these results will have a strong bearing on infection mitigation measures for the hundreds of millions of people driving in passenger cars and taxis worldwide, and potentially yield to safer and lower-risk approaches to personal transportation.

\section*{Methods}

The car geometry was chosen based on the basic exterior of a Toyota Prius. The interior was kept minimal, and comprised of two cylindrical bodies representing the {\it driver} and the {\it passenger}. The CAD model for the car geometry was prepared using SolidWorks, and  subsequent operations including domain discretization (meshing) and case setup were carried out using the ANSYS-Fluent module. 

The steady Reynolds-averaged Navier-Stokes (RANS) equations with a standard $k-\epsilon$ turbulence model was solved on an unstructured grid, made up of about 1 million tetrahedral grid cells. The domain size was $6h \times 5h \times 3h$ in the streamwise, normal, and spanwise directions, respectively, where $h$ is the car height. A single vehicle speed of $U = 22$ m/s (50 mph), which was set as the inflow condition upstream of the front of the car body. A pressure outlet condition was applied at the exit. The simulations were iterated until convergence was achieved for the continuity and momentum equations, and the turbulence dissipation rate, $\epsilon$. Each simulation run took roughly 1.5 hrs of computational time on a standard workstation. A grid-independence study was performed, which established that the resolution adopted was sufficient for the quantities reported in the present work.
 
The mixing and transport of a passive scalar were modeled by solving species transport equations describing an advection-diffusion equation. Separate simulations were performed for the scalar released near {\it driver}, and then for its release near the {\it passenger's} face. The scalar was was set to be a non-interacting material, i.e. with an exceedingly low mass diffusivity, which meant that only advection and turbulent diffusion contributed to its transport dynamics. This approach mimics the mixing of a high Schmidt number material, such as dye or smoke, which are commonly used as a tracers in turbulent fluid flows \cite{almeras2019mixing}. The injection rate of the species was very low in order that it did not influence the air flow. This was verified by comparing the concentration fields for various injection rates, which showed negligible variation. This strategy was followed in order that the effects of turbulent diffusion effects were also captured in the analyses.\\

\small{\noindent \textbf{Acknowledgements:} We thank Siyang Hao and Yuanhang Zhu for useful discussions. We acknowledge the use of images and materials, courtesy of ANSYS, Inc.\\
\noindent\textbf{Funding:} VM acknowledges funding from University of Massachusetts, Amherst startup funds. VM and AD acknowledge funding from the US Army Natick Soldier Systems Center. JAB and KB acknowledge funding from Brown University institutional funds.\\
\noindent \textbf{Author contributions:} KB, JAB and VM conceived the project. VM and KB designed the numerical simulations. AD and VM performed the numerical simulations and data analyses. KB and VM conducted the field experiments. All authors discussed the results and wrote the paper. Email for correspondence: vmathai@umass.edu.}

\bibliography{carflow}


\end{document}


\preprint{AIP/123-QED}

\title{Supplemental  Material \\Airflows inside passenger cars and implications for airborne disease transmission}

\author{Varghese Mathai}
\altaffiliation[]{Email: vmathai@umass.edu}

\affiliation{Department of Physics, University of Massachusetts, Amherst, Massachusetts 01003, USA}
\affiliation{Center for Fluid Mechanics, Brown University, Providence, RI 02912, USA}

\author{Asimanshu Das} 
\thanks{VM and AD contributed equally to this work and are joint first authors.}

\affiliation{Center for Fluid Mechanics, Brown University, Providence, RI 02912, USA}

\author{Jeffrey A. Bailey}
\affiliation{Department of Pathology and Laboratory Medicine, Warren Alpert Medical School, Brown University, Providence, RI 02912, USA}
\author{Kenneth Breuer}
\affiliation{Center for Fluid Mechanics, Brown University, Providence, RI 02912, USA}

\date{\today}

\maketitle

\begin{figure*}[!htbp]
\centering
\includegraphics[width=1\textwidth]{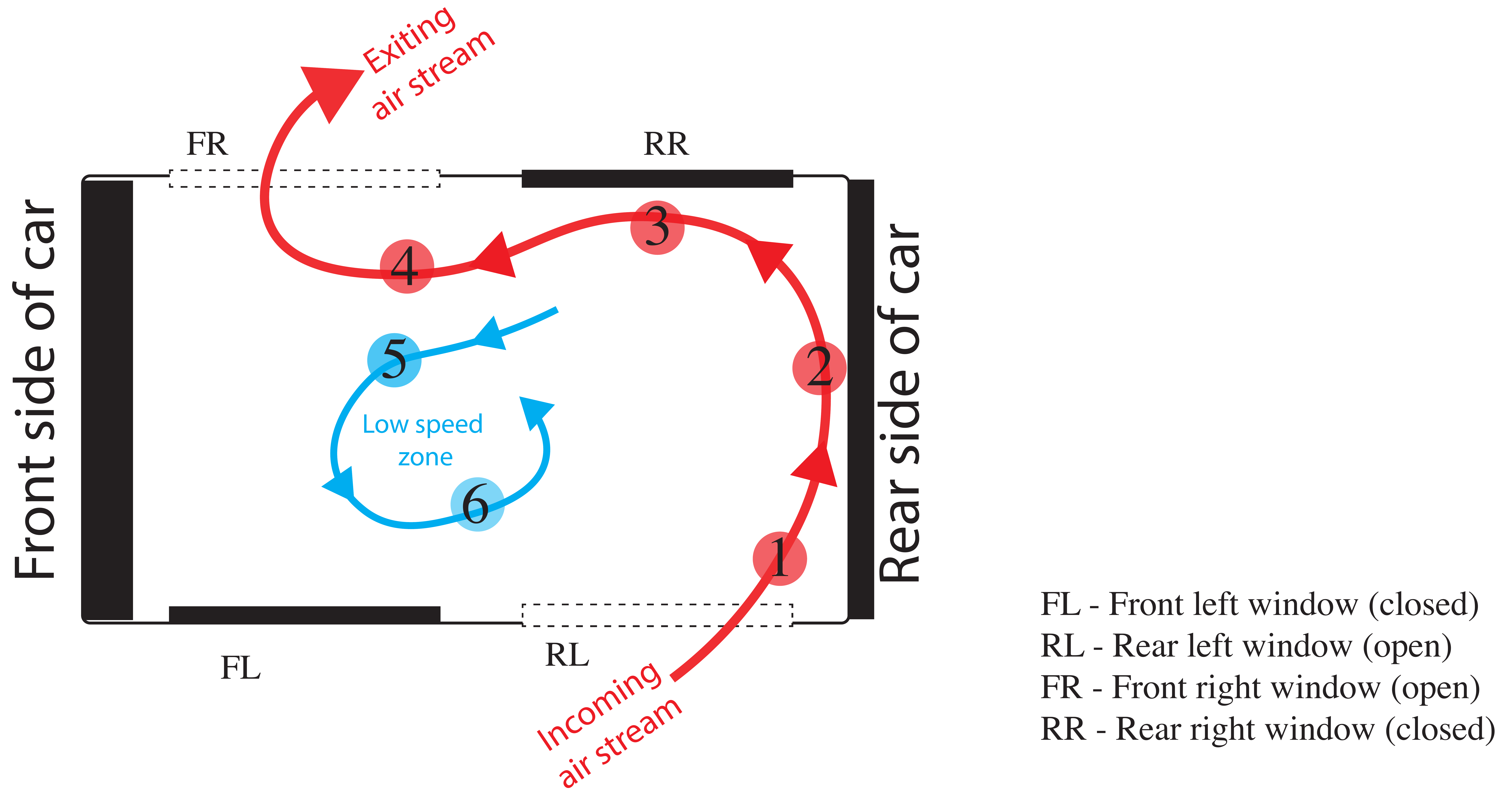}
\caption{Schematic of the approximate flow direction inside the car cabin. The orange streamline illustrates the high speed incoming flow from the rear window, which moves along the back of the car cabin before turning around at the rear left corner and exiting the front window. The blue streamline gives an approximate depiction of the low speed circulation zone in the middle of the cabin. The smoke generator and flow wand techniques in the next figure, substantiate these flow pattern.}
\label{fig:Car_schematpatterns.ic}
\end{figure*}

\begin{figure*}[!htbp]
\centering
\includegraphics[width=1\textwidth]{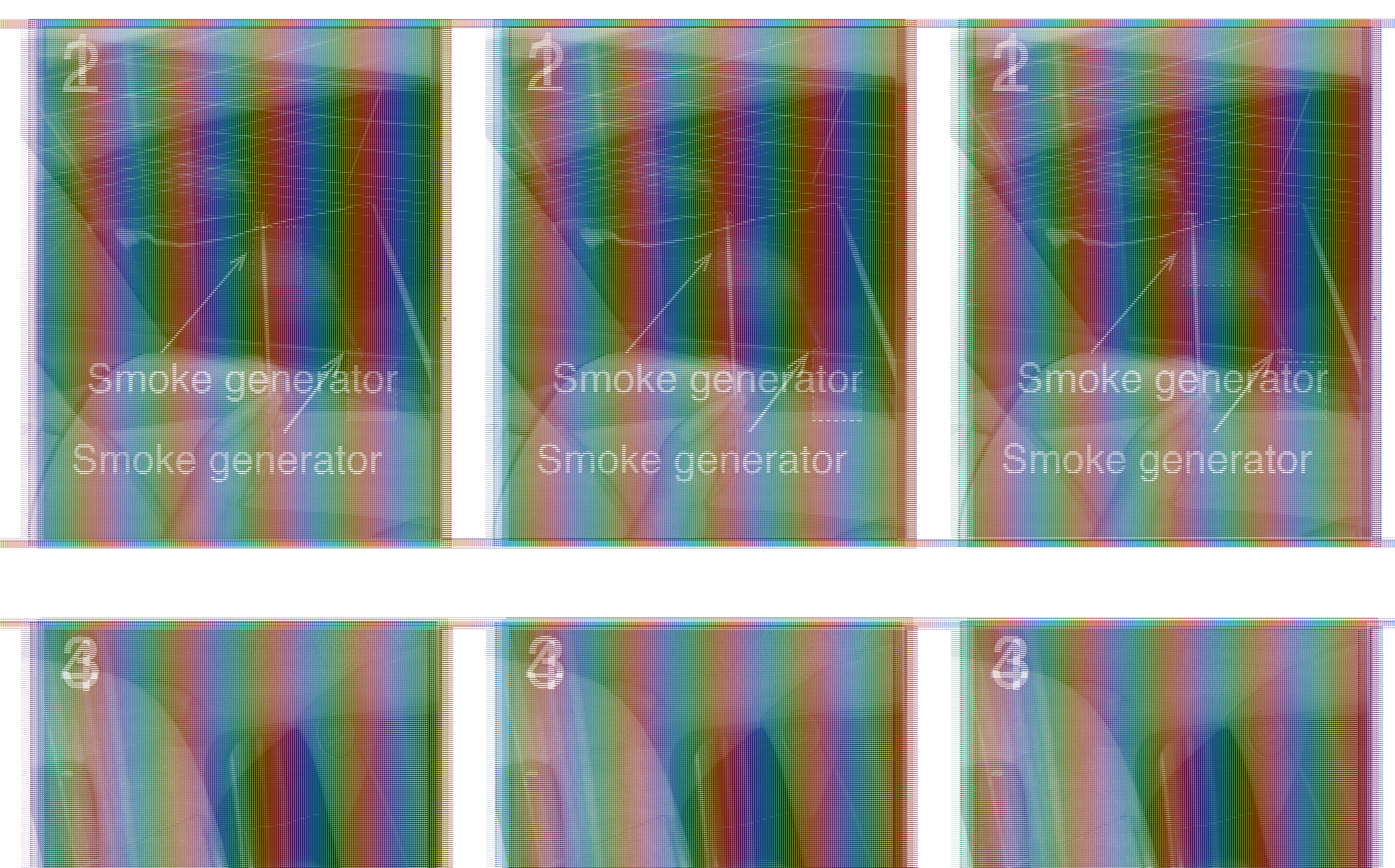}
\caption{Field test images obtained using smoke generator ({\it left sub-figure}) and flow wand ({\it right sub-figure}). The sub-figures (1)-(4) correspond to the region traced by the high speed air stream given in orange in Fig.~S-1. The smoke and flow wand directions are to the left side in the images (1)-(4), which was combined to draw the approximate streamline pattern (orange color) shown in Fig.~S-1. The smoke jet is rapidly pushed in the direction of the flow, which provides a qualitative picture of the high air speed along this streamline path. The field test also confirms the conclusions of the steady RANS simulation discussed in the main paper.}
\label{fig:Main_Streamline}
\end{figure*}

\begin{figure*}[!htbp]
\centering
\includegraphics[width=1\textwidth]{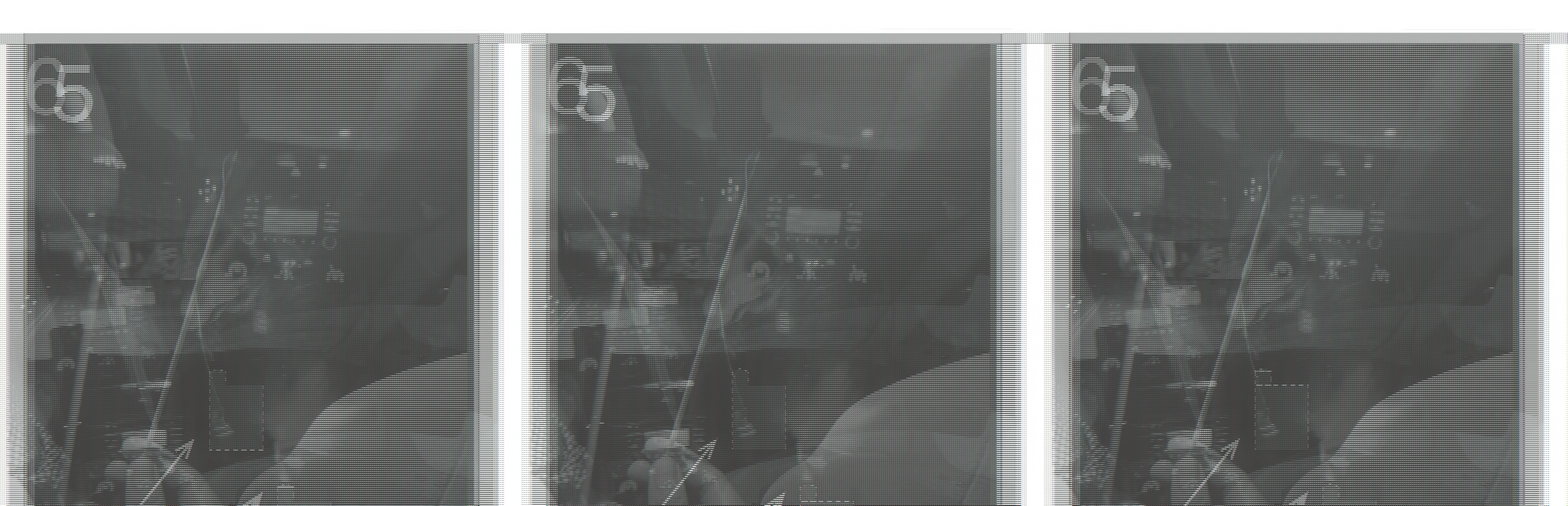}
\caption{Field test images obtained using smoke generator ({\it left sub-figure}) and flow wand ({\it right sub-figure}). The images in sub-figures (5) and (6) were taken near the circulation zone, given in blue in Fig.~S-1. The smoke diffuses very gently in this zone, suggesting the low air flow speed. Similarly, the flow wand fails to catch the wind, and stays almost vertical, suggest a stagnant region of air. This field test provides a validation of the results from the steady RANS simulations.}
\label{fig:Circulation_Streamline}
\end{figure*}

\begin{figure*}[!htbp]
\centering
\includegraphics[width=0.5\textwidth]{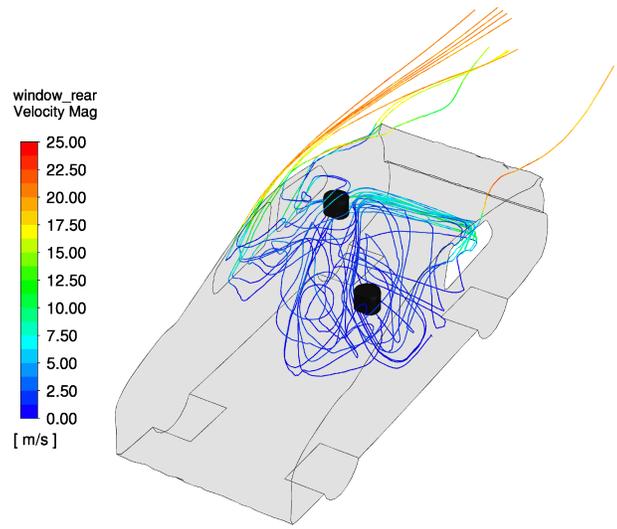}
\caption{Streamlines colored by velocity magnitude for Config. 4 with three windows (FL, RL and RR) open. The opening of the third RR window adjacent to the passenger causes a portion of the incoming air stream to be bled out.}
\label{fig:Pathlines_RL_Config4}
\end{figure*}

\begin{figure*}[!htbp]
\centering
\includegraphics[width=0.6\textwidth]{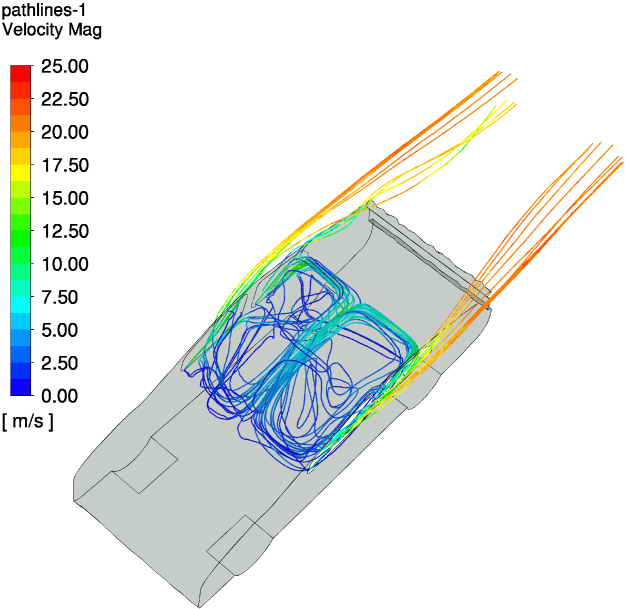}
\caption{Streamlines colored by velocity magnitude for Config. 6 with all four windows open. The flow gets compartmentalized into a left and right zone. In each of these zones the streamlines are directed from the rear window (with higher pressure) toward the front window (with lower pressure).}
\label{fig:Pathlines_Config6}
\end{figure*}

\begin{figure*}[!htbp]
\centering
\includegraphics[width=0.6\textwidth]{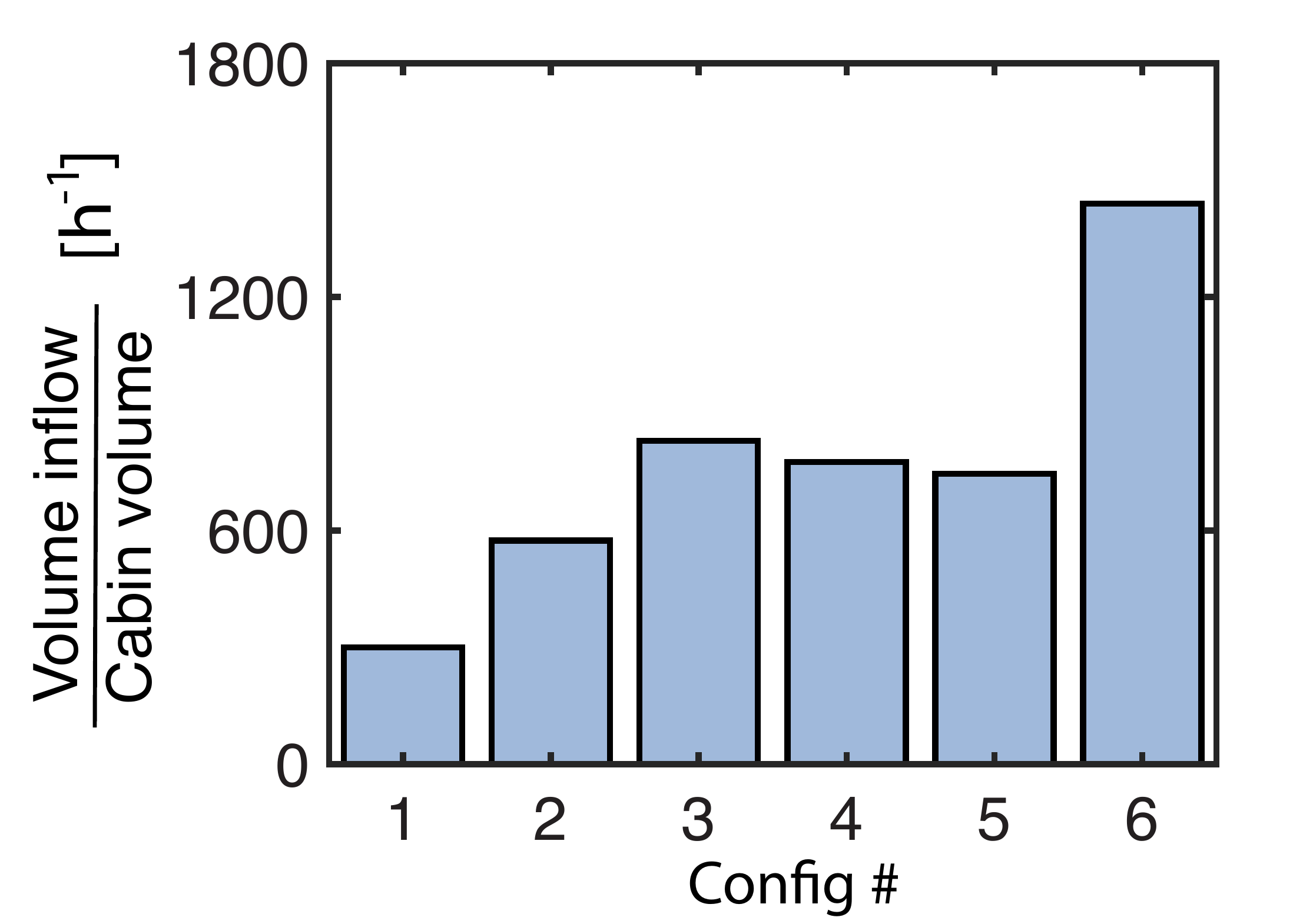}
\caption{Volume of air entering the windows per unit time, compared to the overall cabin volume. The all-windows-open case draws in a significantly more amount of air, whereas the other cases do not exhibit much difference.}
\label{fig:Volumetricflow}
\end{figure*}